\documentclass[12pt]{iopart}

\usepackage[english]{babel}
\usepackage{graphicx}
\usepackage{iopams}
\usepackage{bbm}

\newcommand{\mt}[1]{\mathrm{#1}}
\newcommand{\dd}{\mathrm{d}}
\newcommand{\df}[1]{\mathrm{d}y_{#1}f(y_{#1})}
\newcommand{\dfi}[1]{\mathrm{d}x_{#1}f_{#1}(x_{#1})}
\newcommand{\dfInt}[1]{\mathrm{d} y_{#1}f(y_{#1})\int_{-\infty}^{y_{#1}+c}}

\begin{document}

   \title{Records and sequences of records from
     random variables with a linear trend}
   \author{Jasper Franke, Gregor Wergen and Joachim Krug}
   \address{Institute of Theoretical Physics, University of Cologne,
     Z\"ulpicher Stra\ss e 77, 50937 Cologne, Germany}
   
   \begin{abstract}
     We consider records and sequences of records drawn from discrete
     time series of the form $X_{n}=Y_{n}+cn$, where the $Y_{n}$
     are independent and identically distributed random variables and $c$ is a constant
     drift. For very small and very large drift
     velocities, we investigate the asymptotic behavior of the
     probability $p_{n}\left(c\right)$ of a record occurring in the
     $n$th step and
     the probability $P_{N}\left(c\right)$ that all $N$ entries
     are records, i.e. that $X_1 < X_2 < ... < X_N$. Our work is motivated by the analysis
     of temperature time series in climatology, and by the study of mutational pathways
     in evolutionary biology.
   \end{abstract}

\maketitle

\section{Introduction}

A record is an entry in a discrete time series that is larger
(\textit{upper record}) or smaller (\textit{lower record}) than all
previous entries. In this sense, a record is an extreme value that is
defined relative to all previous values in the time series. 
Record events are of interest in 
various areas of life and science such as 
climatology \cite{Bassett,Benestad,Redner,Meehl,Wergen2} and sports
\cite{Gembris1,Gembris2}, but also in biology \cite{Krug2005,Sire2006,Park2008}. A
record is usually a rare and remarkable event that will be remembered
by observers. Not without good reason the term record originates from
the Latin verb \textit{recordari} - \textit{to recall, to remind}. 

The classic results for records drawn from series of independent and identically
distributed (i.i.d.) random variables (RV's) are well established, 
see \cite{Glick,Nevzorov1987,Arnold,Nevzorov} for review. In this work we
concentrate on two important quantities in particular. The first one
is the probability for a certain entry in a time series to be a record,
and the second one is the probability that the entries of a time
series are ordered, or in other words, that \textit{all} events are
records. For i.i.d. RV's
both these quantities are completely universal for all continuous
probability density functions.
This can be shown by the
so called \textit{stick-shuffling} argument: The last one of $n$
identically distributed entries (sticks) in a time series is equally
likely to be a record as all other entries, and therefore the 
probability $p_n$ for the $n$th event to be a record, henceforth
referred to as the \textit{record rate}, is given by 
\begin{eqnarray}
\label{pnBasic}
p_n = \frac{1}{n}.
\end{eqnarray}
Accordingly the expected mean number of records $R_n$ up to a time
$n$ can be obtained by computing the harmonic sum: $R_n = \sum_{i=1}^n 1/k
\approx \textrm{ln}\left(n\right) + \gamma + O\left(1/n\right)$, where
$\gamma\approx0.577215...$ is the Euler-Mascheroni constant. From
similar considerations one obtains the statistics of waiting times
between record breaking events which turn out to be universal as
well. It is equally straightforward to compute the probability for all events
in a series of length $N$ to be ordered in size. Since this case is
only one of $N!$ possible and equally likely permutations of all $N$
events, the \textit{ordering probability} $P_N$ is given by 
\begin{eqnarray}
\label{PnBasic}
 P_N = \frac{1}{N!}.
\end{eqnarray}
We conclude that the two quantities of interest are related by 
\begin{eqnarray}
\label{independence}
P_N = \prod_{n=1}^N p_n,
\end{eqnarray}
which reflects the fact that record events are independent in the
i.i.d. case \cite{Glick,Arnold}. We will return to this point below in
section 2. In contrast to the properties of record times, 
the distributions of record values are not completely universal, but
their asymptotic behavior falls into three different
universality classes that are analogous to the universality classes of
extreme value statistics: The \textit{Weibull} class of distributions
with finite support, the \textit{Gumbel} class of distributions with
exponential-like tails, and the \textit{Fr\'echet} class of power law tailed
distributions \cite{Galambos,DeHaan,Sornette}. 

Given that the statistics of records for i.i.d. RV's is well understood, it
is natural to ask what happens when the underlying time series is
correlated, or when the RV's are drawn from a distribution that varies
in time. An important example of a correlated random process is the
random walk, and the record statistics of this process was recently
analyzed by Majumdar and Ziff \cite{Majumdar,Majumdar2}. The
simplest realization of a time-dependent distribution is the \textit{linear
drift model} (LDM) first considered by Ballerini and Resnick \cite{Ballerini,Ballerini2}. 
In this model the $n$th entry in the time series is of the form  
\begin{eqnarray}
\label{YnBasic}
 X_n = Y_n + cn,
\end{eqnarray}
where $c$ is a constant and the $Y_n$ are
i.i.d. RV's. In this simple scenario the probability density
$f_n\left(x\right)$ of $X_n$ is of the form
$f_n\left(x\right)=f\left(x-cn\right)$ with a fixed probability density
$f\left(y\right)$ and the corresponding cumulative distribution
function $F(y)=\int_{-\infty}^y dy' \; f(y')$, which is the distribution of the
i.i.d. part $Y_{n}$ of $X_{n}$. We will usually consider upper records
and assume $c > 0$.

The LDM was originally introduced as a model for sports records
in improving populations \cite{Ballerini}, and it has recently appeared
in the context of the dynamics of elastic manifolds in random
media \cite{LeDoussal}. An important motivation for the present work
comes from the interest in the consequences of global warming for the
occurrence of temperature records \cite{Redner,Meehl}.  
In \cite{Wergen2, Wergen} the effect of warming on daily temperature measurements was modeled
using a Gaussian probability density with a linear trend, and it was 
shown that this very simple model is capable of quantitatively 
describing the statistics of record-breaking temperatures at European
and American weather stations. 
In the climate context the drift speed $c$ is typically small compared
to the standard deviation of $f$, which suggests to consider the
behavior of the record rate $p_n(c)$ for small $c$ and finite
$n$. This approach is complementary to previous work on the LDM
\cite{Ballerini,Ballerini2,LeDoussal,Borovkov}, which 
has mostly been concerned with the asymptotic behavior of the record
rate for $n \to \infty$. 

Another application that motivates our research comes from the study
of adaptive paths in evolutionary biology. In this context, a path is
a collection of mutations that change the genotype of an organism into
another genotype of higher fitness.  Given that mutation rates are
small, the evolution of a population usually proceeds
one mutation at a time. For a given set of $N$ mutations,
there are then $N!$ distinct paths which correspond to the different orders
in which the mutations can occur. Since a mutation spreads in
the population only if it confers a fitness advantage, a given
pathway is \textit{accessible} to adaptive evolution only if the
fitness values of the intermediate genotypes increase monotonically
along the path, that is, if they are arranged in ascending order
\cite{Weinreich2005,Weinreich2006}.

In view of the complexity of real fitness landscapes, the intricate
interactions between different mutations are often modeled by
assigning fitness values at random to genotypes \cite{Park2008}. One such model, which
is closely related to the LDM, was introduced by Aita \textit{et al.}
in the context of protein evolution \cite{Aita2000}. In this model the fitness $X_n$ of a
particular intermediate genotype with $n$ mutations is assumed to consist of an i.i.d. RV
$Y_n$ and a systematic part $cn$, where $c > 0$ if the mutations move
the population closer to the global fitness peak, and the value of $c$
(relative to the standard deviation of the $Y_n$) can be adjusted to
tune the ruggedness of the fitness landscape. Taking into account also
the initial genotype with no mutations, a total of $N+1$ genotypes
with fitness values $X_0, X_1,...,X_N$ are encountered along a path.
The probability for a path to be accessible in this model is then just $P_{N+1}(c)$, and the
expected number of accessible paths is $N! P_{N+1}(c)$. An immediate corollary   
of (\ref{PnBasic}) is that the expected number of accessible paths of length
$N$ in a completely random fitness landscape without any average uphill slope ($c = 0$) is $1/(N+1)$ 
\cite{Kloezer2008,FrankeInPrep}. 

Here we consider both the record rate $p_n$ and the ordering
probability $P_N$ for the linear drift model. We 
distinguish between a small drift $c$ that is much smaller than the
characteristic width of the distribution (in most cases the standard
deviation), and a large drift that is much larger than this width. Both
cases are of practical relevance. In section 2 we 
discuss the general properties of record statistics for systems
with linear drift, with particular emphasis on the correlations
between record events. In the subsequent section 3 we will
present new results for small $c$. We examine the
record statistics for members of the three extreme-value classes
individually and find the corresponding asymptotic behaviors.
In section 4 we analyze the case of large $c$. Throughout
Monte-Carlo simulations are used to confirm the analytical
results. Finally, in section 5 we present a
brief summary, discuss related issues and give an outlook on
further possible research. Some of the calculational details are
relegated to Appendices.

\section{General theory and an exactly solvable example}
\label{General}

The values taken by the $\{X_{i}\}_{i\in \{1, \dots, n\}}$ are
stochastically independent. The probability that all $n$ values are
less than a given value $x$ factorizes to
$\prod_{i=1}^{n}\int_{-\infty}^{x} \dd x_{i} f_i\left(x_{i}\right) =
\prod_{i=1}^n F_i(x)$. Here
$f_{i}$ and $F_{i}$ are, as stated in the introduction, the
probability densities and cumulative distribution functions of the
$X_{i}$. Thus, given the value $y_{n}$ of the i.i.d. part $Y_{n}$ of $X_{n}$, the
probability that all previous RV's $\left\{X_{i}\right\}_{i\in \{1,
  \dots, n-1\}}$ are smaller than $X_{n}$ is
$\prod_{i=1}^{n-1}\int_{-\infty}^{y_{n}+ic}\df{n-i}$.
The probability that a RV $X_{n}$ drawn from a general time-dependent distribution $F_{n}(x)$
is a record is therefore given by \cite{Krug2007} 
\begin{eqnarray} \label{p_n_general} \mt{P}\left[X_{n}=\max_{i\in\{1,
      \dots, n\}}\{X_{i}\}\right] = p_n = \int_{-\infty}^\infty \dd x
  f_n\left(x\right) \prod \limits_{i=1}^{n-1}
  F_i\left(x\right),
 \end{eqnarray} 
which reduces to
\begin{equation}\label{Genp}
  p_{n}(c)=
  \int_{-\infty}^{\infty}dx f(x)\prod_{i=1}^{n-1}F(x+ci) 
\end{equation}
for the LDM. It was shown in \cite{Ballerini} that the limiting record rate
\begin{equation}
\label{p(c)}
p(c) \equiv \lim_{n \to \infty}p_{n}(c)= \int_{-\infty}^{\infty}dx f(x)\prod_{i=1}^{\infty}F(x+ci) 
\end{equation}
exists and is nonzero for $c > 0$ provided the distribution $f(y)$ of
the i.i.d. part in (\ref{YnBasic}) has a finite first moment. For $c=0$, 
(\ref{Genp}) can be evaluated directly and,
with the substitution $u=F(x)$, one obtains
\begin{equation}\label{piid}
  p_{n}(c=0)=\int_{-\infty}^{\infty}\dd xf(x)F(x)^{n-1}=
  \int_{F(-\infty)=0}^{F(\infty)=1}\mt{d}uu^{n-1}=\frac{1}{n}, 
\end{equation}
independent of $F$, as already shown in (\ref{pnBasic}).

The other quantity under consideration in this article, the ordering
probability $P_N$, can be expressed as 
\begin{equation}\label{GenP1}
  \fl
  \mt{P}\left[X_{1}<X_{2}<\dots<X_{N}\right]=P_{N}(c)=\int_{-\infty}^{\infty}
  \dfi{N}\dots\int_{-\infty}^{\infty}\dfi{1} \mathbbm{1}_{x_{1}<x_{2}\dots <x_{N}}.
\end{equation}
Inserting $f_{n}(x)=f(x-cn)$ the
indicator function $\mathbbm{1}_{x_{1}<x_{2}<\dots <x_{N}}$ can be
absorbed in the integral boundaries to yield 
\begin{equation}\label{GenP2}
  P_{N}(c)=\int_{-\infty}^{\infty}\dfInt{N}\dd y_{N-1} \dots
  \int_{-\infty}^{y_{2}+c}\df{1}. 
\end{equation}
As for $p_n\left(c\right)$, this equation can be solved for arbitrary
$F$ only in the case $c=0$. Using again the substitution $u=
F(x)$ in turn in all the $N$ integrals, starting from the inside,
one obtains the result already derived in (\ref{PnBasic}), 
\begin{eqnarray}\label{Piid}
  P_{N}(c=0)&=\int_{-\infty}^{\infty}\df{N}\int_{-\infty}^{y_{N}}\dd
  y_{N-1}\dots\int_{-\infty}^{y_{3}}\df{2}F(y_{2})\nonumber \\   
  &=\int_{-\infty}^{\infty}\df{N}\int_{-\infty}^{y}\dd y_{N-1}
  \dots  \int_{-\infty}^{F(y_{3})}\dd uu \nonumber\\
  &=\frac{1}{2}\int_{-\infty}^{\infty}\df{N} \int_{-\infty}^{y_{N}}\dd y_{N-1}
  \dots \int_{-\infty}^{F(y_{4})}\mt{d}uu^{2}=\dots=\frac{1}{N!}.  
\end{eqnarray}
The reason for re-deriving the two previous results is that here this
is done in a way that in principle generalizes to arbitrary $c$. 

For $c>0$, the exact evaluation of Eqs.(\ref{Genp}) and (\ref{GenP2}) has
proven difficult, but in the case where the $Y_{n}$ are Gumbel
distributed, i.e. $F(y)=\exp\left(-e^{-y}\right)$, one can use the
fact that this distribution obeys the relation
$F(y+a)=F(y)^{\exp(-a)}$ to 
explicitly perform the integration in (\ref{Genp}) \cite{Ballerini,Ballerini2}. 
With the abbreviation $\alpha\equiv e^{-c}$ and the substitution
$u=F(y)$ one obtains  
\begin{equation}\label{Gumbelp}
  p_{n}(c)=\int_{-\infty}^{\infty}\dd yf(y)F(y)^{\sum_{i=1}^{n-1}\alpha^{i}}=
  \left(\sum_{i=0}^{n-1}\alpha^{i}\right)^{-1}=\frac{1-e^{-c}}{1-e^{-nc}}
\end{equation}
by use of the incomplete geometric series. Keeping $c$
fixed, one obtains in the limit $n \to \infty$ the asymptotic record
rate $p(c)=1-e^{-c}$,
while for $c \to 0$ one recovers the i.i.d. result $p_n = 1/n$.
For $c < 0$ the record rate is seen to decay exponentially in $n$,
which implies that the expected number of records $R_n$ remains finite
for $n \to \infty$. We suspect this to be a general feature of the LDM
with $c < 0$, but are not aware of a proof of this fact.
 
The relation used to evaluate (\ref{Gumbelp}) for the Gumbel case 
can also be used in (\ref{GenP2}), as before starting from the innermost
integral, which yields
\begin{eqnarray}\label{GumbelP}
    P_{N}(c) &=\int_{-\infty}^{\infty}\dfInt{N}\dd y_{N-1}\dots
  \int_{-\infty}^{y_{3}+c}\df{2}F(y_{2}+c)\nonumber\\
  &=\int_{-\infty}^{\infty}\dfInt{N}\dd y_{N-1}\dots
\int_{-\infty}^{F(x_{3}+c)}\mt{d}uu^{\alpha} \nonumber\\
 &=\frac{1}{\alpha
    +1}\int_{-\infty}^{\infty}\df{N}\int_{-\infty}^{x_{N}+c}\dd y_{N-1}
  \dots\int_{-\infty} 
  ^{F(x_{4}+c)}\mt{d}u u^{\alpha 
    (\alpha+1)}\nonumber\\
  &=\dots =\prod_{l=1}^{N-1}\frac{1}{\sum_{k=0}^{l}\alpha^{k}} .
\end{eqnarray} 
Summing the geometric series as in (\ref{Gumbelp}), one obtains
\begin{equation}\label{GumbelPartition}
  P_{N}(c)=\left(1-e^{-c}\right)^{N}\frac{1}{\prod_{n=1}^{N}\left(1-e^{-cn}\right)}\equiv\left(1-e^{-c}\right)^{N}\mathcal{Z}_{N},
\end{equation} 
where $\mathcal{Z}_{N}$ is the grand canonical partition function
of a system of bosonic particles with energy levels $n
= 1,...,N$ at inverse temperature $c$. This partition function also occurs as one limit in the integer partition problem
(see \cite{Comtet1, Comtet2} and references therein).

\begin{figure}
  \includegraphics[scale=0.6]{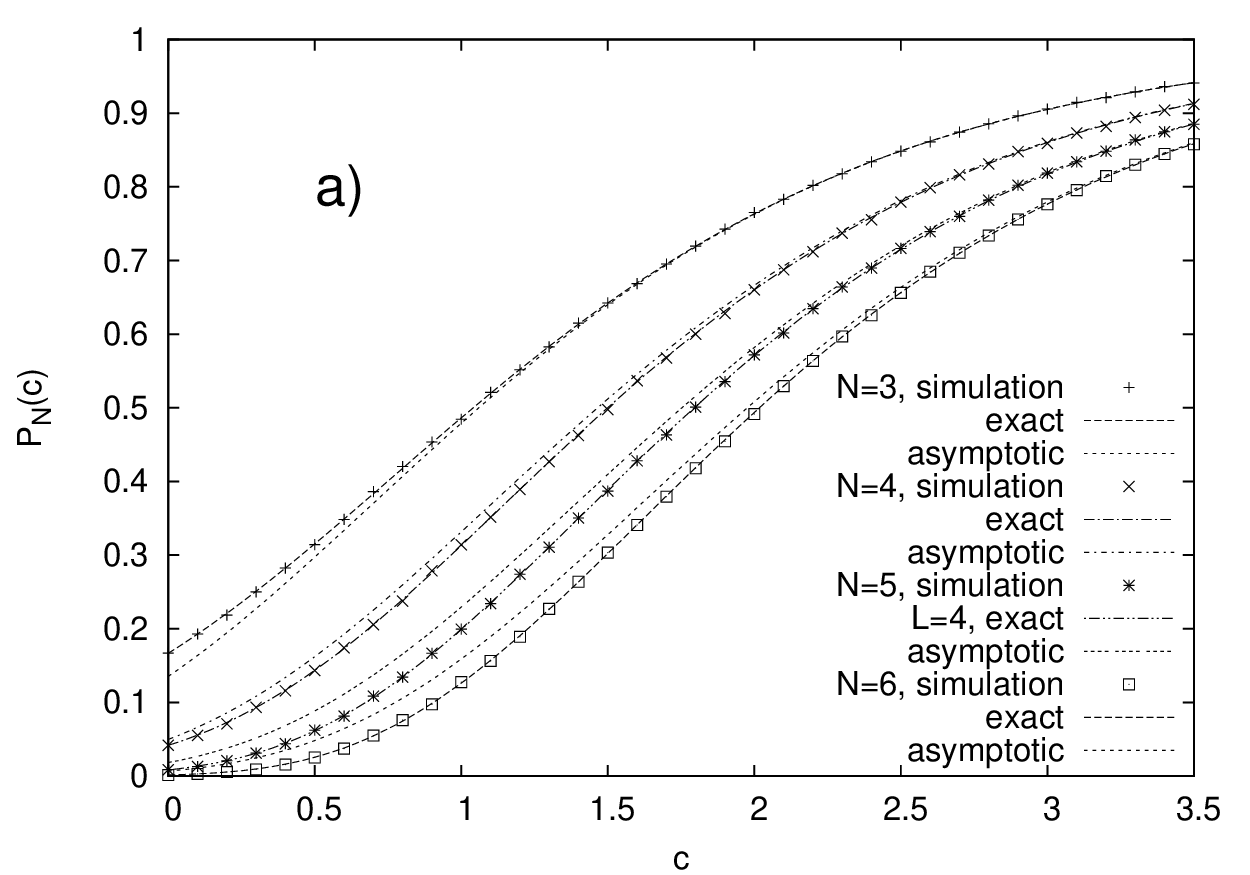}\hfill
  \includegraphics[scale=0.6]{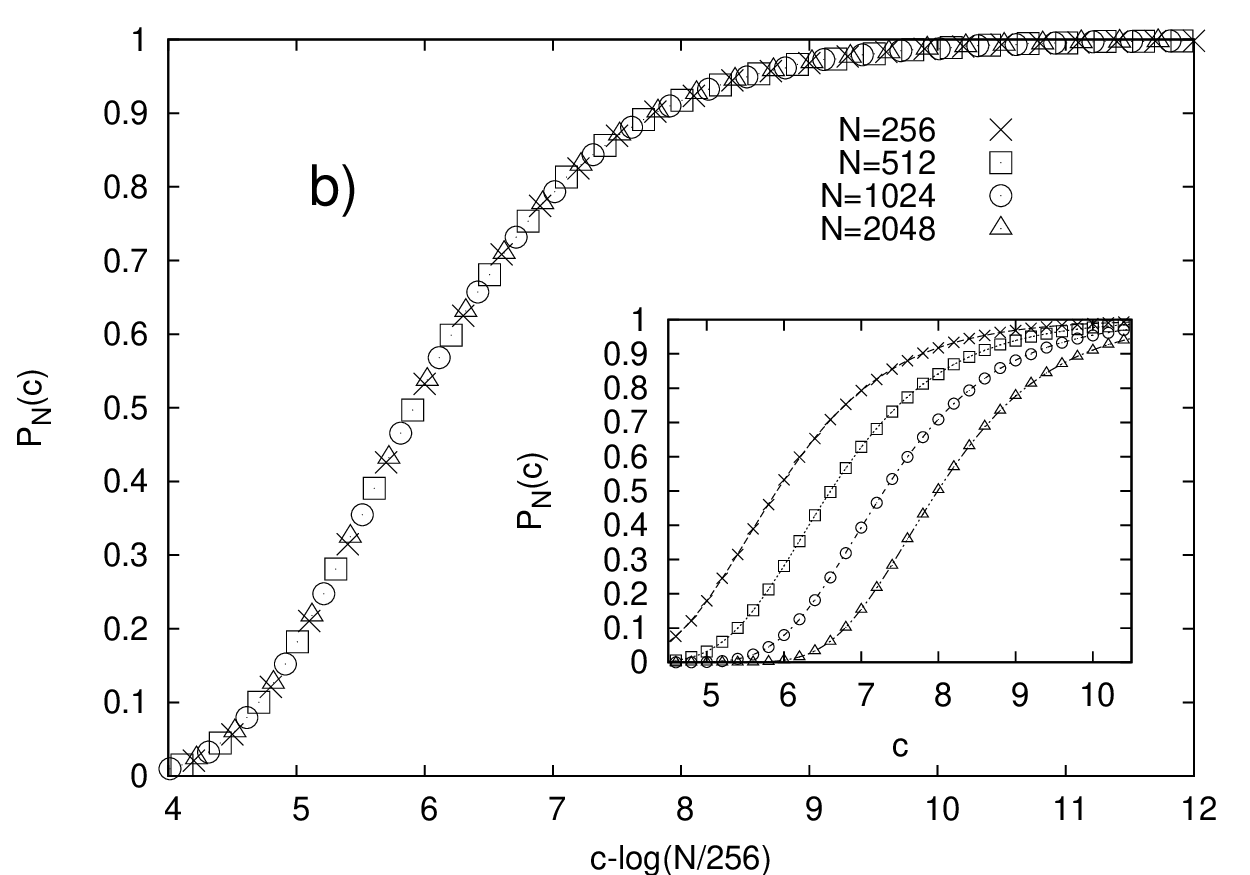}
  \caption{\label{GumbelSimulation} Comparison of the exact
    expression (\ref{GumbelP}) and the asymptotic
    expression (\ref{GumbelApp}) to numerical simulation. The exact
    expression is confirmed, and while there is a
    clear difference between simulations and asymptotic expression for
    small values of $c$ in \textbf{a)}, the approximation holds with
    good accuracy for 
    large $c$ (inset of \textbf{b)}, lines are the asymptotic
    expressions). The main plot of \textbf{b)} demonstrates 
    the scaling between $N$ and $c$ according to (\ref{GumbelApp}).}
\end{figure}

The product $\prod_{n=1}^{N}\left(1-\exp(-cn)\right)$ in the
denominator is the so-called $q$-Pochhammer symbol $(q;q)_{N}$ with $q
= e^{-c}$. In the limit $N \to \infty$ with fixed $c$, one has the
asymptotic expression \cite{wolfram2}
\begin{equation}
\lim_{N \to \infty} \prod_{n=1}^{N}(1-e^{-cn}) \equiv (e^{-c})_\infty
\approx  
\sqrt{\frac{2 \pi}{c}}\exp\left(-\frac{\pi}{6c}+\frac{c}{24}\right),
\end{equation}
and thus, by inserting this into (\ref{GumbelPartition}), 
\begin{equation}
  P_{N}(c)\approx
  \sqrt{\frac{c}{2\pi}}\exp\left(N \ln (1 - e^{-c}) +\frac{\pi}{6c}-\frac{c}{24}\right)
  , \mbox{\hspace{8pt}} N\gg 1. 
\end{equation}
On the other hand, taking $c \gg 1$ at fixed $N$, one has $\alpha =
\exp(-c) \ll 1$ and thus the geometric series in the denominator 
of (\ref{GumbelP}) can be approximated to first order in
$\alpha\equiv\exp(-c)$, as $1/(\sum_{k=0}^{l}\alpha^{k})\approx 1-\alpha
+\mathcal{O}(\alpha^2)$. Then (\ref{GumbelP}) becomes 
\begin{equation}\label{GumbelApp}
  P_{N}(c) \approx
  \exp\left(-(N-1)\alpha\right)=\exp\left(-(N-1)e^{-c}\right),\mbox{\hspace{8pt}} c \gg 1.
\end{equation}
This expression is distinguishable from numerical data only in the
region of $c \sim \mathcal{O}(1)$, see figure \ref{GumbelSimulation}. 

Comparing the exact expressions Eqs.(\ref{GumbelP}) and
(\ref{Gumbelp}), one sees that the relation (\ref{independence})
obtained in the i.i.d. case remains valid here. This is a consequence of
the mutual stochastic independence of record events in the LDM with
Gumbel-distributed i.i.d. part \cite{Nevzorov,Nevzorov1987,Borovkov}.
In fact the Gumbel distribution is uniquely characterized by the
mutual independence of record values and record indicator variables (which 
indicate whether or not a record occurs at time $n$) \cite{Borovkov,Ballerini1987}.

\begin{figure}[t]
  \centerline{\includegraphics[scale=0.8]{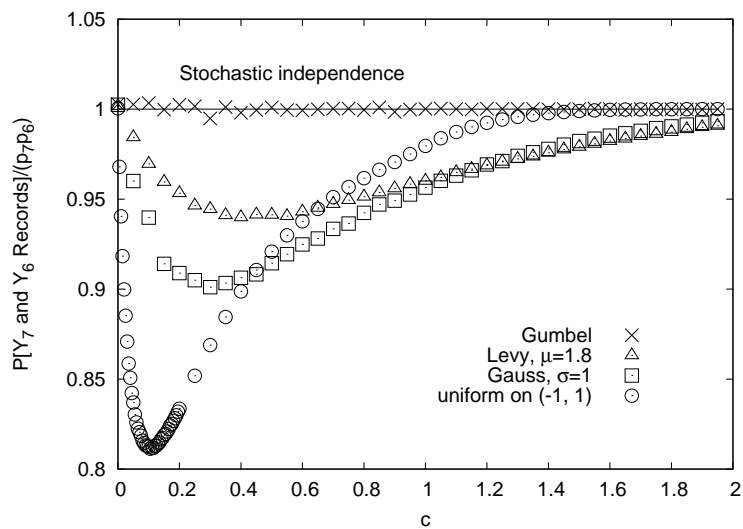}}
  \caption{\label{CorRecEv} Joint probability of two consecutive
    record events at times $n=6$ and 7, divided by the product of the corresponding record
    rates. This ratio is unity if record events are stochastically independent. For $c=0$, this is the case, just as
    for Gumbel-distributed i.i.d. parts (crosses). Note that for other
    probability densities, the record events also seem to become increasingly
    independent as $c$ grows.} 
\end{figure}

For $c>0$ and arbitrary distribution $F$, however, the record events
in the LDM are not independent. Numerical 
studies for several different distributions presented in
figure \ref{CorRecEv} show that the records are negatively
correlated and seem to repel each other. A more
thorough examination of the structure of correlations between record
events in this model is currently ongoing research \cite{InPrep}.
For the purpose of the present discussion we merely note that record
events appear to become asymptotically
uncorrelated for large $c$. This fact will be used to derive some asymptotic
results for $P_{N}(c)$ in section \ref{largec}. First however we consider the case $c\ll 1$.


\section{Record statistics for small drift}
\subsection{Record rate}

In the previous section we gave a general expression for the record
rate $p_n\left(c\right)$ of the LDM. 
Here, we derive the first order term in a series expansion for $c\ll 1$. If $c$
is very small (\ref{Genp}) can be simplified as follows: 
\begin{eqnarray} p_{n}\left(c\right) & = & \int_{-\infty}^\infty \dd y
  f\left(y\right) \prod \limits_{i=1}^{n-1} F\left(y+ci\right) \nonumber
\\ & \approx & \int_{-\infty}^\infty \dd y f\left(y\right) \prod
\limits_{i=1}^{n-1} \left[ F\left(y\right) + ci f\left(y\right)\right] \nonumber
\\ & \approx & \int_{-\infty}^\infty \dd y f\left(y\right)
F^{n-1}\left(y\right) +
c\frac{n\left(n-1\right)}{2}\int_{-\infty}^\infty \dd y f^2\left(y\right)
F^{n-2}\left(y\right) \nonumber \\
 \label{p_nc_general} & = & \frac{1}{n} + c I_n
\end{eqnarray}
with 
\begin{equation}
\label{In}
I_n \equiv \frac{n\left(n-1\right)}{2}\int_{-\infty}^\infty \dd y
 f^2\left(y\right) F^{n-2}\left(y\right). 
\end{equation}
This expansion is valid provided $f\left(y\right)$ is slowly varying
between $y$ and $y+ci$, which strictly speaking requires $nc$ to be small compared to
the width of the distribution. In the following we will evaluate the
first order correction coefficient $I_n$ for several
elementary distributions. 

Before doing this, we show that our formula for $p_{n}\left(c\right)$ can be
generalized with respect to the position of the record in the
time-series. Specifically, we consider the probability
that the $k$th event in a time-series of length $n$ with linear drift
$c$ is a record. For this purpose we have to consider the following
integral instead of (\ref{p_n_general}): 
\begin{eqnarray} \mt{P}[X_k =
    \textrm{max}\left(X_1,...,X_n\right)] =
  \int_{-\infty}^\infty \dd y f_n\left(y\right) \prod \limits_{i=1,i\neq
    k}^{n-1} \int_{-\infty}^{y+c\left(i-k\right)} \dd y_i
  f_i\left(y_i\right). \end{eqnarray} 
Evaluating this integral in the same way as shown above, we
obtain the following expression: 
\begin{eqnarray} \mt{P}[X_k =
    \textrm{max}\left(X_1,...,X_n\right)] & \approx &
  \frac{1}{n} +
  \frac{c}{2}\left(k^2-k-\left(n-k\right)\left(n-k-1\right)\right)\times \nonumber
  \\  
& & \times\int_{-\infty}^\infty \dd y f^2\left(y\right) F^{n-2}\left(y\right).
\end{eqnarray}
Note that for $k=n$ this expression reduces to our approximation (\ref{p_nc_general}) for
$p_{n}\left(c\right)$. Apparently for $c>0$ this expression
assumes its maximum for $k=n$ and its minimum for $k=1$. The last
entry has the largest, and the first entry the smallest chance to
be the maximum of the series. 

\subsubsection{Weibull class.}
Let us start by considering the Weibull class of extreme value
statistics, which contains distributions with finite support. 
A simple example for a member of
the Weibull class is a uniform distribution, which takes the value
$\frac{1}{2a}$ between $-a$ and $a$ and $0$ outside of this
interval. For this case the first order expansion of $p_n\left(c\right)$ is given by
\begin{eqnarray}
p_{n}^{uniform}\left(c\right) = \frac{1}{n} +
c\frac{n\left(n-1\right)}{2}\int_{-\infty}^\infty \dd y
\left(\frac{1}{2a}\right)^2\left(\frac{y}{2a} +
  \frac{1}{2}\right)^{n-1}  + O\left(c^2\right), 
\end{eqnarray}
which can be evaluated to yield
\begin{eqnarray}
p_{n}^{uniform}\left(c\right) \approx \frac{1}{n} + c\frac{n-1}{4a}.
\end{eqnarray}
\noindent In this case the correction coefficient $I_n$ increases
linearly with the number of events $n$.

More generally, we consider distributions of the form
\begin{equation}
\label{Weibull}
f\left(y\right) = \xi\left(1-y\right)^{\xi-1} 
\end{equation}
with $\xi>0$ and $0<y\le1$. For these distributions we have 
\begin{eqnarray}
 p_{n}\left(c\right) \approx \frac{1}{n} + c\frac{n\left(n-1\right)}{2}\int_0^1 \dd y
 \xi^2 \left(1-y\right)^{2\xi-2}
 \left(1-\left(1-y\right)^\xi\right)^{n-2}. 
\end{eqnarray}
The integral is divergent for $\xi < 1/2$, which indicates that
$p_n(c)$ is a non-analytic function of $c$; this case will be
considered elsewhere. For $\xi > 1/2$ 
we use the substitution $\left(1-y\right) = z^{1/\xi}$ to express the
integral in terms of a Beta-function,  
\begin{eqnarray}
 p_{n}\left(c\right) \approx \frac{1}{n} +
 c\xi\frac{n\left(n-1\right)}{2}\frac{\Gamma\left(2-\frac{1}{\xi}\right)\Gamma
   \left(n-1\right)}{\Gamma\left(n+1-\frac{1}{\xi}\right)}.
\end{eqnarray}
Using the Stirling approximation for large $n$ one finally arrives at 
\begin{eqnarray}
\label{In_Weibull}
 p_{n}\left(c\right) \approx \frac{1}{n} +
 \frac{c\xi}{2}\Gamma\left(2-\frac{1}{\xi}\right)n^{\frac{1}{\xi}},
\end{eqnarray}
which shows that $I_n$ generally increases as a power law in the
Weibull class. 

\subsubsection{Fr\'echet class.}
\begin{figure}\begin{center}\includegraphics[width=0.9\textwidth]{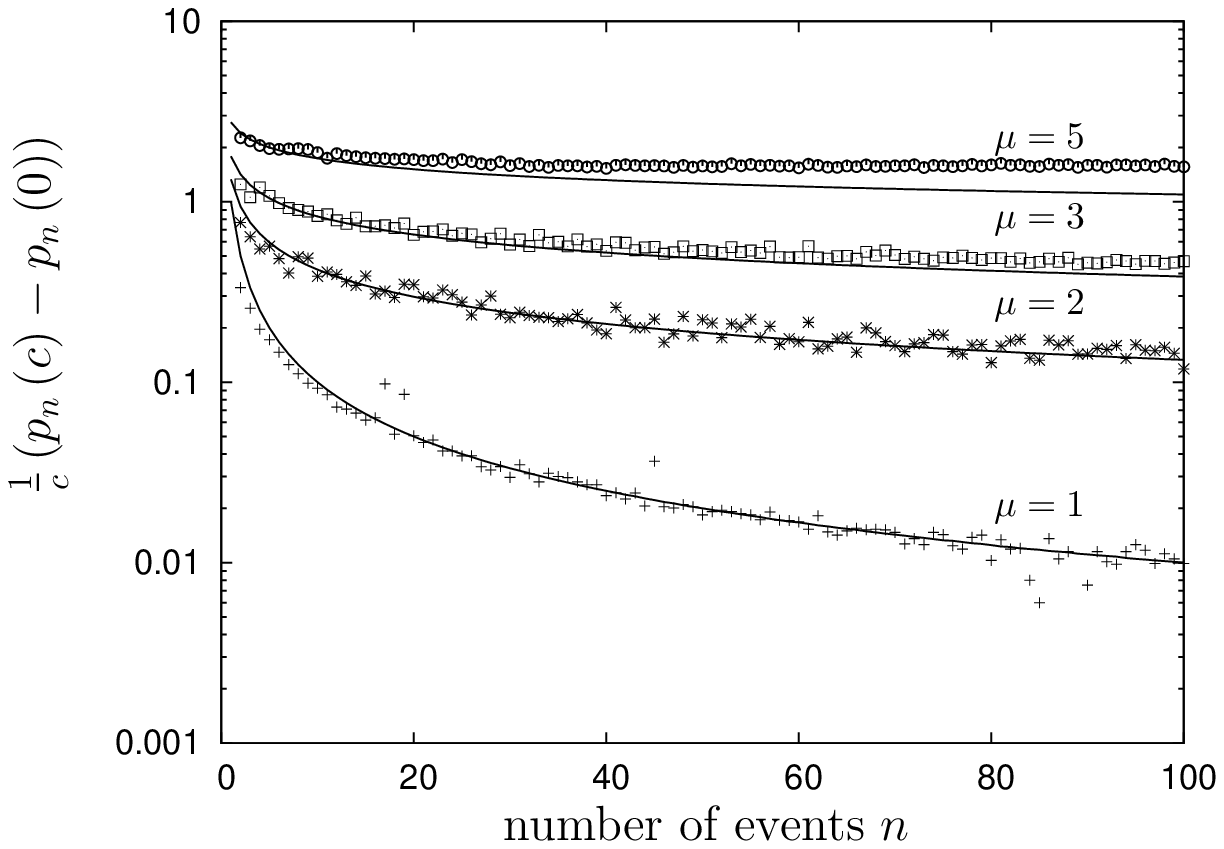}
\caption{Results of Monte-Carlo simulations of the LDM for power law tailed distributions of the Fr\'echet
      class. The figure shows the difference between the record
      rate in the time-independent case for $c=0$ and the
      drifting case with drift $c=0.01$. This difference is
      given by $\frac{1}{c}\left(p_{n}\left(c\right)-p_{n}\left(0\right)\right)$. The dots
      correspond to simulations with different tail coefficients $\mu=1,2,3, 5$ averaged over $10^6$ runs, and the lines
      show the analytic predictions. The first order approximation is very good for $\mu=1$ and $\mu=2$
      but becomes less accurate for larger $\mu$.\label{FrechetFigure}}\end{center}\end{figure}

As a representative of the Fr\'echet class of extreme value statistics we
consider a general power-law distribution of the form
$f\left(x\right)=\left(1/\mu\right) x^{-\mu-1}$ for $x>1$ and
$\mu>0$. For distributions of this kind $p_{n}\left(c\right)$ in the
small $c$ expansion is given by
\begin{eqnarray}
 p_{n}\left(c\right) \approx \frac{1}{n} +
 c\frac{n\left(n-1\right)}{2}\int_1^\infty \dd y \mu^2
 y^{-2-2\mu}\left(1-y^{-\mu}\right)^{n-2}. 
\end{eqnarray}
Again, the integral is very similar to a Beta-function and it can be
transformed into one by elementary means. Doing this we find 
\begin{eqnarray}
 p_{n}\left(c\right) \approx \frac{1}{n} +
 c\mu\frac{n\left(n-1\right)}{2}\frac{\Gamma\left(2+\frac{1}{\mu}\right)\Gamma\left(n-1\right)}{\Gamma\left(n+\frac{1}{\mu}+1\right)},  
\end{eqnarray}
and using again the Stirling approximation we obtain
\begin{eqnarray}
\label{In_Frechet}
p_{n}\left(c\right) \approx \frac{1}{n} +
\frac{c\mu}{2}\Gamma\left(2+\frac{1}{\mu}\right)\frac{1}{n^{1/\mu}}. 
\end{eqnarray}
In figure \ref{FrechetFigure} we compare this prediction to simulation results.

While in the case of the Weibull class the correction term $I_n$
increases with $n$, here it decays as a power-law
$n^{-1/\mu}$. For $\mu > 1$ the decay is slower than the $1/n$-decay
of the record rate in the absence of a drift, which implies that the
drift will nevertheless dominate the behavior for long times. This is
consistent with the fact that the record rate reaches a nonzero
asymptotic limit, as given by (\ref{p(c)}), because $\mu > 1$ 
implies a finite first moment for the $Y_n$. On the other hand, for 
$\mu < 1$ the decay of $I_n$ is faster $1/n$ and the limit on the
right hand side of (\ref{p(c)}) vanishes for any $c$, which implies
that the drift is asymptotically irrelevant. The borderline situation $\mu
= 1$ has been studied by De Haan and Verkade \cite{DeHaan1987}, who
find that the asymptotics depends nontrivially on the value of $c$ in this case. 

In general the results presented so far show that the effect of the drift on a broad
distribution is smaller than on a more narrow distribution. 
A similar qualitative trend was found in \cite{Krug2007}
for probability densities with increasing variance.

\subsubsection{Gumbel class.}
The Gumbel class comprises unbounded distributions that decay
faster than any power-law. A very simple representative of the
Gumbel class is the exponential distribution $f\left(y\right) =
\nu^{-1} e^{-\frac{y}{\nu}}$. In this case the first order expansion
(\ref{p_nc_general}) assumes the following form: 
\begin{eqnarray}
 p_{n}^{exp}\left(c\right) \approx \frac{1}{n} + c\frac{n\left(n-1\right)}{2}\int_0^{\infty}\dd y\frac{1}{\nu^2}e^{-\frac{2y}{\nu}}\left(1-e^{-\frac{y}{\nu}}\right)^{n-2}.
\end{eqnarray}
The integral can be solved by two partial integrations and one finds
\begin{eqnarray}
 p_{n}^{exp}\left(c\right) \approx \frac{1}{n} + \frac{c}{2\nu},
\end{eqnarray}
that is, the correction term is independent of $n$.

The calculation for the Gaussian distribution, arguably the most
important member of the Gumbel class, is more complicated. For convenience we consider a Gaussian
distribution of unit variance,
\begin{eqnarray} f\left(y\right) =
  \frac{1}{\sqrt{2\pi}}e^{-\frac{y^2}{2}}. \end{eqnarray}
The integral of interest reads
\begin{eqnarray}
  I_n^{gauss} = 
  \frac{n\left(n-1\right)}{2\sqrt{2\pi}^n} \int_{-\infty}^\infty \dd y
  e^{-y^2}\left(\int_{-\infty}^y \dd {y'} e^{-\frac{-{y'}^2}{2}}\right)^{n-2}, 
\end{eqnarray}
which will be evaluated for large $n$ using the saddle point approximation. 
With the definition 
\begin{equation}
\label{g}
g\left(y\right) := -y^2 +
\left(n-2\right)\ln\left(\frac{1}{\sqrt{2\pi}}\int_{-\infty}^y \dd {y'}
  e^{-\frac{{y'}^2}{2}}\right)
\end{equation}
we have
\begin{eqnarray}
\label{saddle_point_eq}
 I_{n}^{gauss} \approx \frac{n\left(n-1\right)}{4\pi} \sqrt{\frac{-2\pi}{\dd_y^2 g\left(\tilde{y}\right)}}e^{g\left(\tilde{y}\right)},
\end{eqnarray}
where $\tilde{y}$ denotes the saddle-point of the integral. It turns out
that the computation of a practicable series-expansion of
$g\left(y\right)$ can only be done under some approximations and by
using the non-elementary Lambert-W
function \cite{Knuth,Wolfram}. In terms of the 
W-function $\textrm{W}\left(z\right)$ defined by the relation $\textrm{W}(z) e^{\textrm{W}(z)} = z$, we find 
\begin{eqnarray}
 \tilde{y} = \sqrt{\textrm{W}\left(\frac{\left(n-2\right)^2}{8\pi}\right)}.
\end{eqnarray}
For large $z$ the Lambert-W function can be approximated by
$\textrm{W}\left(z\right) \approx \textrm{ln}\left(z\right) -
\textrm{ln}\left(\textrm{ln}\left(z\right)\right)$, which eventually yields 
\begin{eqnarray}
\label{Gauss}
p_{n}^{gauss}\left(c\right) \approx \frac{1}{n} +
c\frac{2\sqrt{\pi}}{e^2}\sqrt{\textrm{ln}\left(\frac{n^2}{8\pi}\right)}.
\end{eqnarray}
For a detailed derivation of this result see \textbf{APPENDIX I}. 
In figure \ref{GaussFigure} the asymptotic prediction is compared to numerical
simulations. The systematic deviations that are visible in this figure
can be attributed to strong sub-leading corrections to
(\ref{Gauss}), see \textbf{APPENDIX I}. 

\begin{figure}\begin{center}\includegraphics[width=0.8\textwidth]{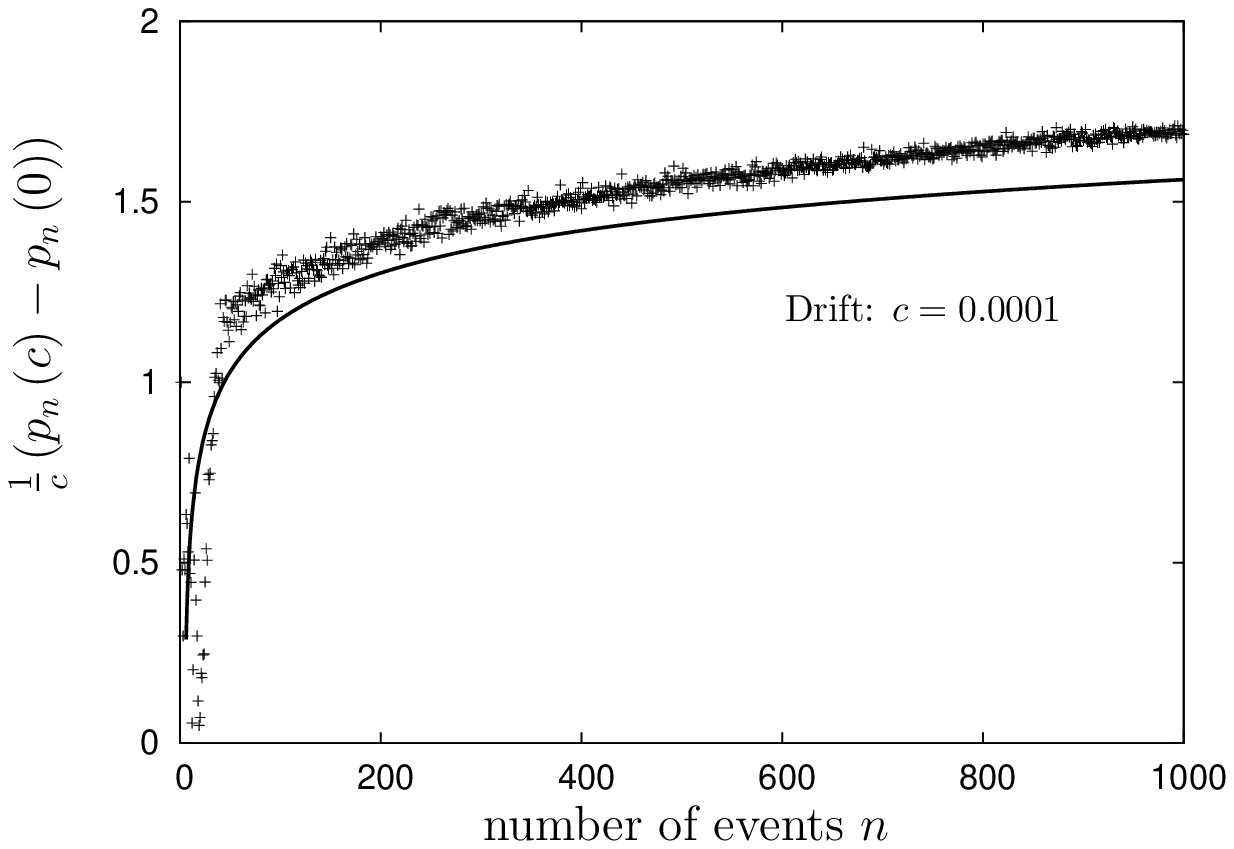}\caption{Results 
      of Monte-Carlo simulations from $10^9$ realizations of the LBM with RV's
      drawn from a normal distribution with standard deviation
      $\sigma=1$. The figure shows the normalized difference between the
      record rate in the time-independent case and the drifting
      case, $\frac{1}{c}\left(p_{n}\left(c\right)-p_{n}\left(0\right)\right)$.  The dots correspond to a simulation
      with drift velocity $c=10^{-4}$.\label{GaussFigure}}\end{center}\end{figure} 

As a more general subset of the Gumbel class we also
considered distributions of the form $f\left(y\right) = C_\beta e^{-|y|^\beta}$ with
$\beta>0$ and normalization constant $C_\beta = [2 \Gamma(1 + 1/\beta]^{-1}$. The integral of interest
then reads 
\begin{eqnarray}
\label{beta_integral}
I_n = \frac{n\left(n-1\right)}{2} C_\beta^n\int_{-\infty}^\infty \mathrm{d}y
 e^{-2|y|^{\beta}}\left(\int_{-\infty}^y \mathrm{d}y' e^{-|y'|^\beta}
 \right)^{n-2}, 
\end{eqnarray}
which can again be treated using a saddle-point
approximation. Ignoring constant prefactors we find that
\begin{eqnarray}
\label{In_Gumbel}
I_n \propto \textrm{ln}\left(n\right)^{1-\frac{1}{\beta}} 
\end{eqnarray}
for large $n$, which includes the results for the exponential distribution ($\beta
= 1$) and the Gaussian ($\beta = 2$) as special cases. 
For a detailed derivation of this result see \textbf{APPENDIX II}. 
We conclude that the behavior of the correction coefficient $I_n$ in the Gumbel
class is generally intermediate between the power law growth for
distributions in the Weibull class, and the power law decay for
Fr\'echet-type distributions. Again, the effect of the drift is stronger for distributions that fall off more rapidly 
(large $\beta$).

\subsubsection{Relation to the asymptotic record rate $p(c)$.}
It is instructive to compare the asymptotics of the correction term
$I_n$ derived in the preceding subsections to the behavior of the 
limiting record rate $p(c)$ for small $c$, which was studied
by Le Doussal and Wiese \cite{LeDoussal}. Heuristically, the two
quantities can be related as follows. We have seen above that, for any
choice of $f(y)$ with a finite first moment, 
the correction term $I_n$ becomes large compared to
$1/n$ for large $n$. This implies that, for any $c > 0$, the first order
correction will eventually become comparable to the zero'th order record
rate $1/n$. The corresponding time scale $n^\ast$ can be estimated from 
\begin{equation}
\label{nast1}
n^\ast I_{n^\ast} \sim c.
\end{equation}
For times $n > n^\ast$ the first-order expansion breaks down and the
record rate saturates at a nonzero limiting value $p(c)$. Thus we expect
that, in order of magnitude, $p(c) \sim 1/n^\ast(c)$. Using the
asymptotic results (\ref{In_Weibull},\ref{In_Frechet},\ref{In_Gumbel})
together with (\ref{nast1}) we may then determine the behavior of
$p(c)$ for small $c$. The result
\begin{equation}
p(c) \sim  \left\{ \begin{array}{l@{\quad \quad}l}
c^{\xi/(1+\xi)} & \textrm{Weibull} \\
c^{\mu/(\mu-1)} & \textrm{Fr\'{e}chet} \; \textrm{with} \; \mu > 1 \\
c \vert \ln c \vert^{1 - 1/\beta} & \textrm{Gumbel}
\end{array} \right.
\end{equation}
agrees with the analysis of \cite{LeDoussal} in all cases.

\subsection{Ordering probability}


In this subsection, we derive a first order expansion for the
ordering probability $P_N(c)$. Our main result reads 
\begin{equation}\label{GenSmallc}
  P_{N}(c) = \frac{1}{N!}
  +c\frac{1}{(N-2)!}\int_{-\infty}^{\infty}\mt{d}x f^2(x)+\mathcal{O}(c^2).
\end{equation}
In contrast to the expansion (\ref{p_nc_general}) for the record rate,
one sees that for $P_{N}(c)$ the distribution $f(x)$ only enters
in the form of a non-universal constant but has no
influence on the $N$-dependence of the correction term. Note, however,
that similar to the expansion for $p_n(c)$, the correction term
diverges when $f^2(x)$ becomes too singular, as is the case for the
Weibull-type distribution (\ref{Weibull}) with $\xi < 1/2$.

\begin{figure}
  \includegraphics[scale=0.55]{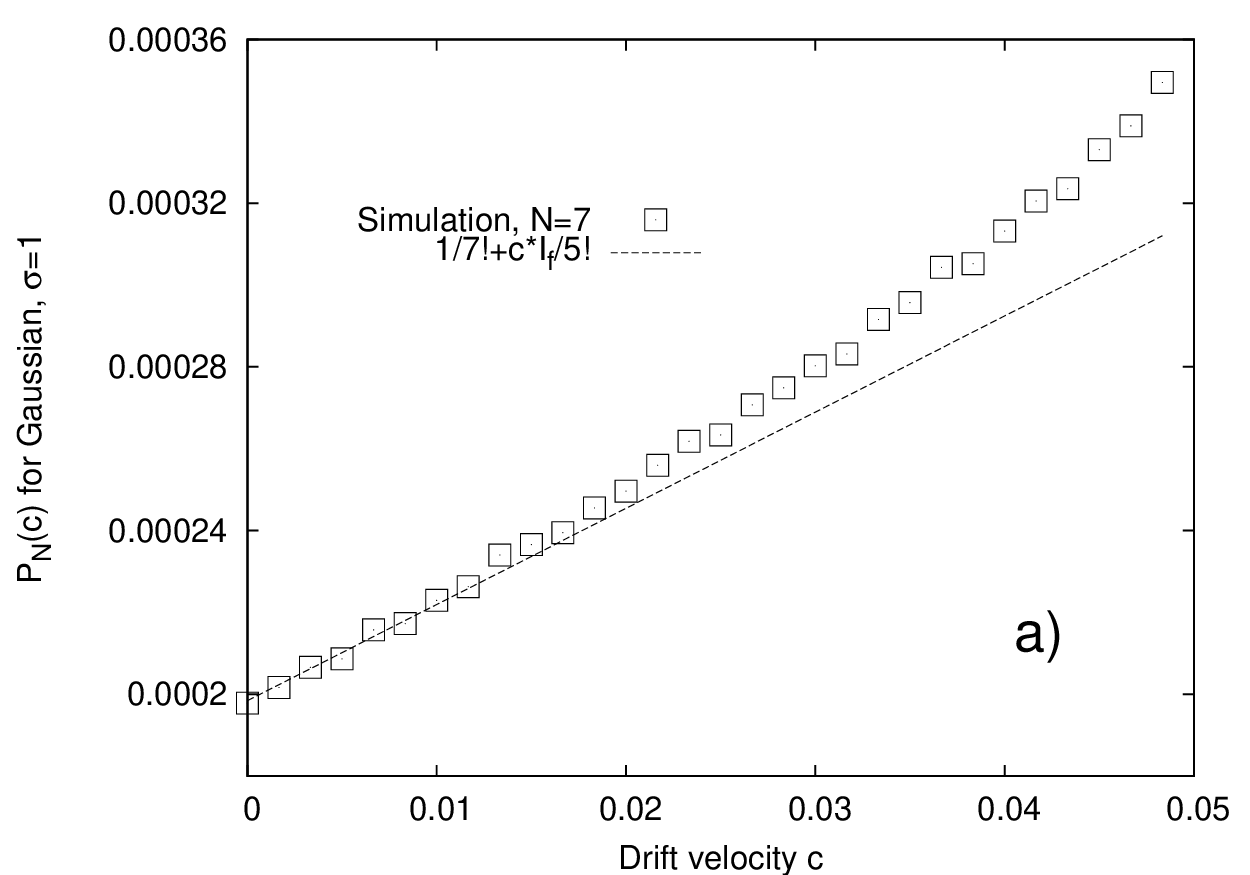}\hfill
  \includegraphics[scale=0.55]{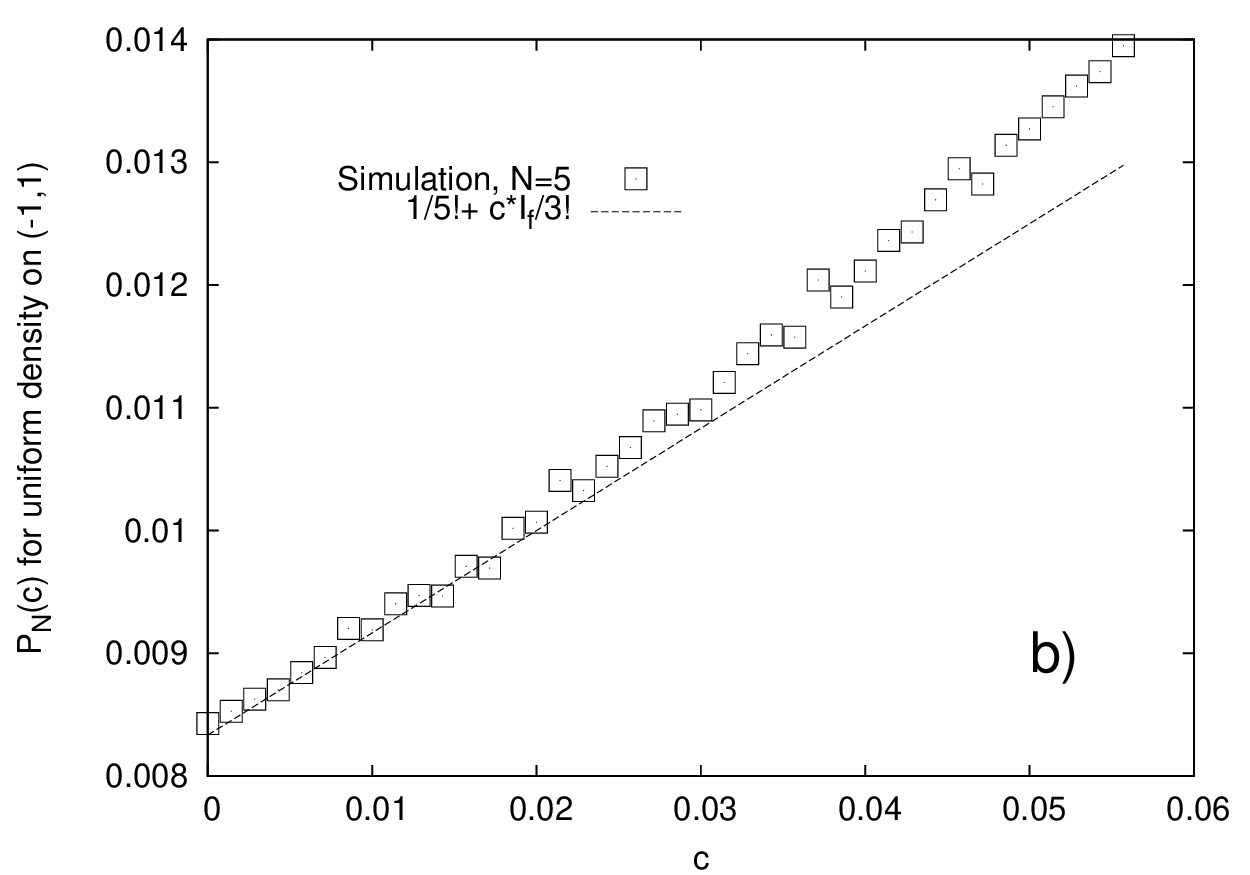}
  \caption{\label{CompareSmallC} Simulations comparing the ordering probability
    $P_{N}(c)$ to the first order expansion $P_N(c) = 1/N! + c I_f/(N-2)!$, where $I_f = \int dy \; f(y)^2$, for  \textbf{a)}
    Gaussian distribution and $N=7$, \textbf{b)} uniform distribution
    and $N=5$.}
\end{figure}

To prove (\ref{GenSmallc}), we set up a Taylor expansion of (\ref{GenP2})
in $c$ to first order. With $P_{N}(0)=1/N!$ we have
\begin{equation*}
  \fl
  P_{N}(c)= \frac{1}{N!}+c\left.\frac{\mt{d}}{\mt{d}
      c}P_{N}(c)\right|_{c=0}+\mathcal{O}(c^2) \approx \frac{1}{N!}+\int_{-\infty}^
  {\infty}\mt{d}x_N f(x_{N})\left.\frac{\mt{d}}{\mt{d}
      c} P(N-1,c,x_{N})\right|_{c=0}, 
\end{equation*}
where terms of $\mathcal{O}(c^2)$ and higher have been omitted and 
\begin{eqnarray*}
  \fl P(N-1, c, x_{N})&\equiv
  \int_{-\infty}^{x_{N}+c}\dfInt{N-1}\dd y_{N-2}\dots
   \int_{-\infty}^{y_{2}+c}\df{1}\\
 \fl &=\int_{-\infty}^{x_{N}+c}\df{N-1} P(N-2,c, y_{N-1}).
\end{eqnarray*}
Clearly, the derivative of $P(N-1, c, x_{N})$ obeys the recursion
relation
\begin{eqnarray*}
  \left.\frac{\dd}{\dd c}P(N-1,c,
    x_{N})\right|_{c=0}&=&f(x_{N})P(N-2, 0,
  x_{N})\\ 
  &&+\int_{-\infty}^{x_{N}+c}\df{N-1}\left.\frac{\dd}{\dd c}
    P(N-2,c, y_{N-1})\right|_{c=0}.
\end{eqnarray*}
Using the same substitutions as in (\ref{Piid}), one obtains
\begin{equation*}
  P(N-2,c=0, x_{N})=\frac{1}{(N-2)!}F^{N-2}(x_{N})
\end{equation*}
and thus
\begin{eqnarray*}
  \left.\frac{\dd}{\dd c}P(N-1, c,
    x_{N})\right|_{c=0}=&\frac{f(x_{N})}{(N-2)!}F^{N-2}(x_{N})\\ 
  &+\int_{-\infty}^{x_{N}+c}\df{N-1}\left.\frac{\dd}{\dd c}
    P(N-2,c, y_{N-1})\right|_{c=0}. 
\end{eqnarray*}
Now $P(1, c, x_{2})=\int_{-\infty}^{x_{2}+c}\df{1}$ and thus
$\left. \frac{\dd}{\dd c}P(1, c, x_{2})\right|_{c=0}=f(x_{2})$. Putting
this into the recursion relation above and integrating over all
$y_{N}$ weighted by $f(y_{N})$, we obtain
\begin{eqnarray}\label{ChainOfIntegrals}
\fl  \left.\frac{\dd}{\dd c}P_{N}(c)\right|_{c=0}=&\frac{1}{(N-2)!}
  \int_{-\infty}^{\infty}\dd y_{N}f^{2}(y_{N})F^{N-2}(y_{N})\nonumber\\
  \fl &+\frac{1}{(N-3)!}\int_{-\infty}^{\infty}\df{N}\int_{-\infty}^{y_{N}}
 \dd y_{N-1}f^{2}(y_{N-1})F^{N-3}(y_{N-1}) +\dots \nonumber \\
  \fl &+\frac{1}{0!}\int_{-\infty}^{\infty}\df{N}\int_{-\infty}^{y_{N}}
 \df{N-1}\dots\int_{-\infty}^{y_{2}}\dd y_{1} f^2(y_{1}),
\end{eqnarray}
a sum with $N$ terms, the last of which comprises $N-1$ nested
integrals. Somewhat miraculously, as shown in \textbf{APPENDIX III}, this chain of integrals can be
collapsed into the simple closed form advertised in (\ref{GenSmallc}).
Figure \ref{CompareSmallC} compares the asymptotic
expression for $P_{N}(c)$ derived here with numerical simulations.


\section{Record statistics for large drift}\label{largec}


\begin{figure}
  \includegraphics[scale=0.55]{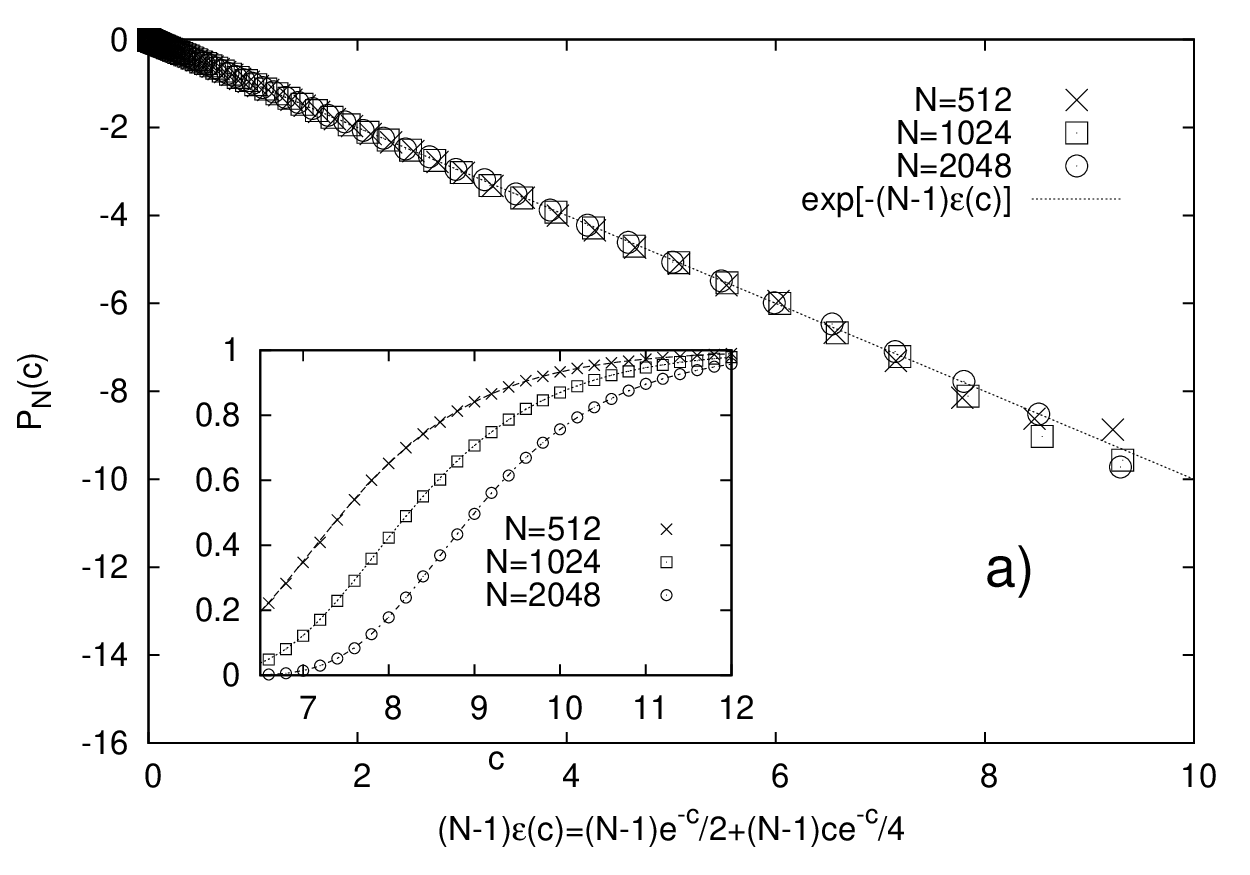}\hfil
  \includegraphics[scale=0.55]{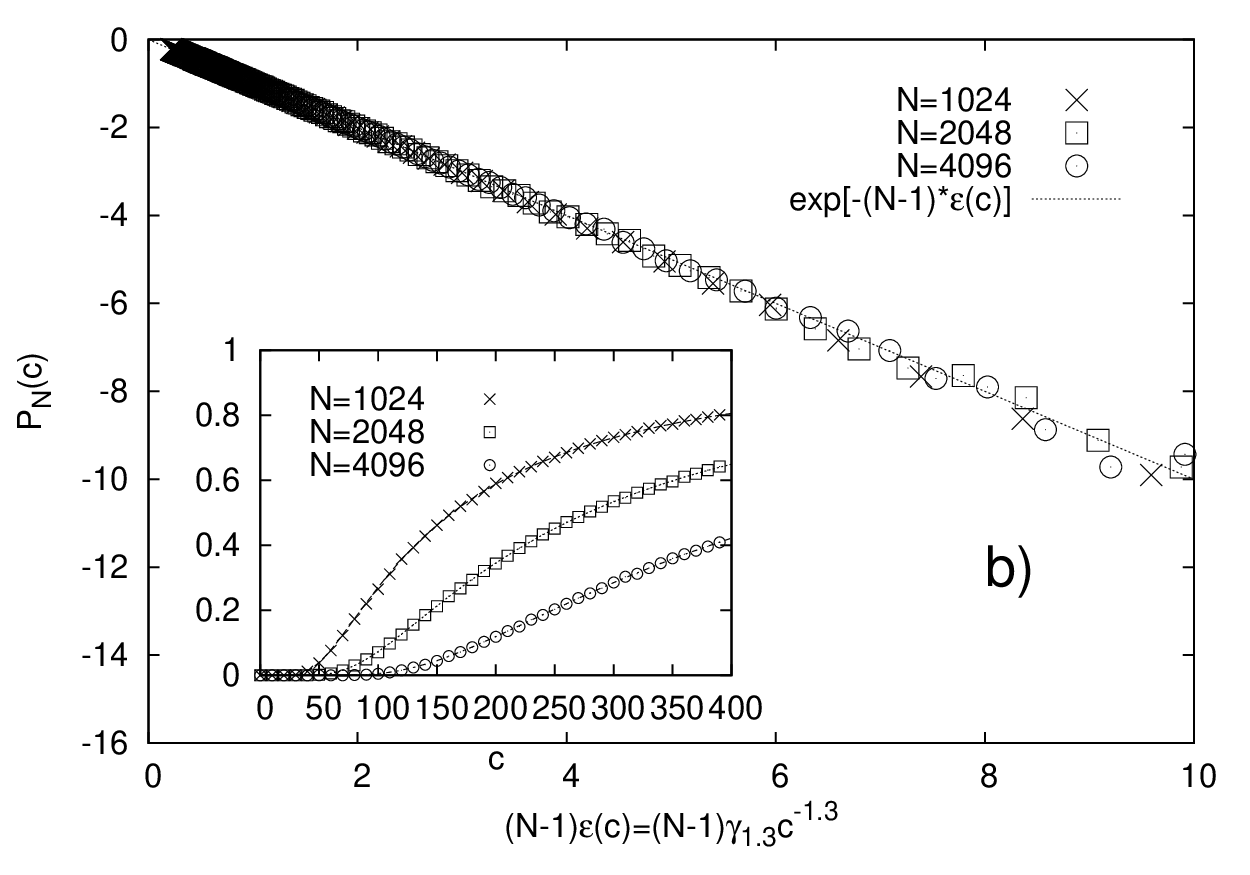}
  \caption{\label{NumericalCollapse} Scaling collapse of $P_N(c)$ as suggested
    by the asymptotic expression (\ref{AsymptoticPL}) for
    \textbf{a)} Laplace density $f(x)=e^{-|x|}/2$ and \textbf{b)} L\'evy-density with
    $\mu=1.3$. The 
    ordinate is the corresponding expression from 
    Eqs.(\ref{epsiLa}) and (\ref{epsiLe}) respectively. Note that the
    asymptotic expressions get more accurate for larger $N$. Inset shows
    direct plots of simulation results (points) versus asymptotic expression (lines).}
\end{figure}

In section \ref{General} we saw that, although record events in the LDM are 
generally correlated for $c>0$, the correlations tend to diminish for
large $c$ (figure \ref{CorRecEv}). This is in some sense expected, as 
for $c \to \infty$ both $p_n(c)$ and $P_N(c)$ tend to unity, such that
the stochastic independence relation (\ref{independence}) becomes
trivially satisfied. Moreover, numerical studies \cite{Wergen} suggest that
the rate of convergence of the record rate to its limiting value
$p(c)$ increases with $c$ and for
sufficiently large values is to a good accuracy attained from the
very beginning. Thus for large $c$ (\ref{independence}) can be approximated
by 
\begin{equation}\label{AsymptoticPL}
  P_{N}(c) \approx p(c)^{N}=\left(1-\epsilon(c)\right)^{N-1}\approx
  e^{-(N-1)\epsilon(c)},   
\end{equation}
where $\epsilon(c)$ is the probability that $X_{n}$ is
\emph{not} a record. For large $c$, only $X_{n-1}$ has an
appreciable chance of keeping $X_{n}$ from being a record. Thus 
\begin{equation}\label{GeneralEpsilon}
  \epsilon(c)\approx
  \mathrm{P}\left[X_{n-1}>X_{n}\right]=\int_{c}^{\infty}\mathrm{d}xf^{*2}(x). 
\end{equation}
Here $f^{*2}(x)$ denotes the twofold convolution of the 
probability density $f(x)$ of the i.i.d. part of $X_{n}$.
To quote a few examples:
\begin{eqnarray}
  \fl
  f(x)=\frac{1}{\sqrt{2\pi}}e^{-x^2/2} \Rightarrow
  \epsilon(c)=\frac{1}{2}\mathrm{erfc}(c/2)\approx \frac{1}{c\sqrt{\pi}}e^{-c^2/4} \label{epsiGa}\\
  \fl
  f(x)=\frac{1}{2}e^{-|x|} \Rightarrow
  \epsilon(c)=\frac{1}{2}e^{-c}+\frac{c}{4}e^{-c}\label{epsiLa}\\
  \fl
  f(x)=\frac{1}{2\pi}\int_{-\infty}^{\infty}e^{-ikx+|k|^{\mu}}\Rightarrow
  \epsilon(c)=\frac{1}{2\pi}\int_{c}^{\infty}\mathrm{d}x\int_{-\infty}^{\infty}
  \mathrm{d}ke^{-ikx-2|k|^{\mu}}\approx
  \gamma_{\mu}
  c^{-\mu},\label{epsiLe}
\end{eqnarray}
with
\begin{equation}
\label{gamma}
\gamma_{\mu}=\frac{2\Gamma(1+\mu)\sin\left(\frac{1}{2}\pi\mu\right)}{\pi \mu}.
\end{equation} 
The first two of these examples are from the Gumbel class of extreme value
statistics, whereas the third example is from the Fr\'echet class
\cite{Galambos, DeHaan}. The asymptotic expression in (\ref{epsiGa}) is from
\cite{Stegun}, while the one in (\ref{epsiLe}) can straightforwardly be
derived from the known expression for the large-$x$ asymptotics for
$f(x)$, see e.g. \cite{Bouchaud1990}. Note that the large $c$ asymptotics for
the  Weibull class is trivial, because both $p_n$ and $P_N$ become
identically equal to unity once $c$ exceeds the range of support of
$f(y)$. Inserting the expressions (\ref{epsiGa},\ref{epsiLa},\ref{epsiLe}) into (\ref{AsymptoticPL}) and also
considering the asymptotics of the exact expression for $P_{N}(c)$
derived in (\ref{GumbelApp}), we see
that in the limit of large $N$ and $c$ the behavior of 
the ordering probability is generally of the approximate form
\begin{equation}
\label{PNlarge}
P_N(c) \approx \exp[-N/N^\ast(c)],
\end{equation}
where $N^\ast(c) \sim e^c$ for the Gumbel and exponential
distributions, $N^\ast(c) \sim e^{c^2}$ for the Gaussian, and 
$N^\ast(c) \sim c^\mu$ for the L\'evy distribution.

\begin{figure}[h]
 \centerline{\includegraphics[scale=0.7]{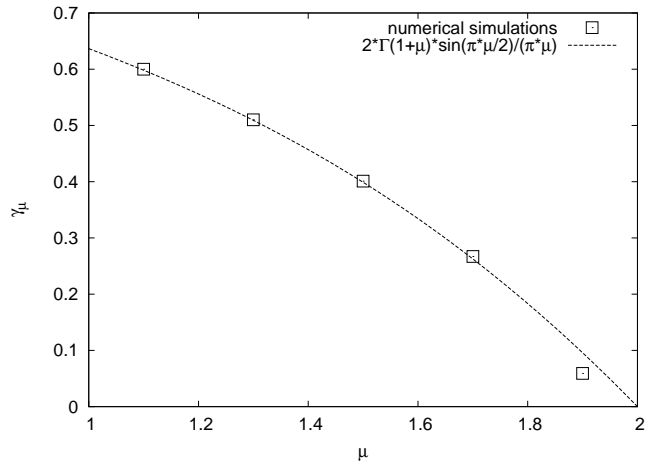}}
  \caption{\label{Gammacheck} Check of the expression for
    $\gamma_{\mu}$ from (\ref{epsiLe}). For $N=1024$, the range
    $0\leq c \leq 400$ was numerically explored as for the data shown in
    figure \ref{NumericalCollapse}. The curves obtained this way were
    then fitted to the form
    $\exp\left(-N\gamma_{\mu}c^{-\mu}\right)$. The value of
    $\gamma_{\mu}$ obtained in this way is shown here for various
    values of $\mu$ and compared to the analytic expression
    (\ref{gamma}).}
\end{figure}

To verify the approximations made in this section, we performed numerical simulations, see
figure \ref{NumericalCollapse} and figure \ref{Gammacheck}. The results indicate that
our approach, although quite rough and not necessarily
well-controlled, does indeed capture the interesting regime rather well
for sufficiently large $N$ and $c$.


\section{Conclusions}

In this article we considered the statistics of records and sequences
of records of random variables with a linear trend as described by
(\ref{YnBasic}). We numerically explored the 
correlations between record events (cf. figure \ref{CorRecEv}) and
analytically investigated the record rate $p_{n}(c)$ and the ordering probability 
$P_{N}(c)$
in the limiting regimes of small and large drift velocities, $c \ll 1$
and $c \gg 1$ respectively. For the regime of $c\sim\mathcal{O}(1)$,
we have not found a generally applicable method. Thus the behavior of $p_{n}(c)$
and $P_{N}(c)$ in this regime remains an open problem.

Specifically, 
we considered the effect of a small linear drift on distributions of the three extreme-value classes. 
While this effect is varying even within the individual classes we still found systematic
differences between them. For the Fr\'echet class of distributions with
power-law tails we found that the coefficient of the leading order correction to the record rate
decays as $I_n \sim n^{-1/\mu}$ for large $n$. This implies a
distinction between distributions with and without a finite first
moment: For $\mu > 1$ the correction decays more slowly than the
unperturbed record rate $1/n$, which implies that the drift dominates
asymptotically and $p_n(c)$ attains a nonzero limit for $n \to
\infty$; on the other hand, for $\mu < 1$ the drift is asymptotically
irrelevant. 

For the considered
distributions of the Gumbel class the situation was a bit more
complicated. For the exponential distribution we found a constant
additive correction to the record rate, while
for generalized Gaussian probability densities $f\propto e^{-|x|^\beta}$ 
the correction term was shown to be of order
$\textrm{ln}\left(n\right)^{1-\frac{1}{\beta}}$, which increases (decreases) with $n$ when $\beta > 1$ ($\beta < 1$). 
For the distributions of
the Weibull class, the effect of the drift is the strongest, and the
correction term generally increases as a power law in $n$. Moreover,
for highly singular distributions with $\xi < 1/2$ in (\ref{Weibull}),
we found indications for a non-analytic behavior of $p_n(c)$ which
will be investigated elsewhere.  Generally speaking, narrow distributions are very sensitive to drift,
while for broad distributions with heavy tails the effect is much
weaker. We have also pointed out that the behavior of the
first order correction term $I_n$ obtained in this paper precisely
matches earlier results for the asymptotic record rate $p(c)$
\cite{LeDoussal}. 

For the probability of a sequence of $N$ consecutive records, we find
the following: For $c \ll 1$, the distribution $f(y)$ of the i.i.d. part of
$X_{n}$ enters to leading order in $c$ only as a numerical constant
$\int_{-\infty}^{\infty}\mathrm{d}xf^{2}(x)$, see (\ref{GenSmallc}), but
the $N$-dependence is 
completely universal for all distributions for which the integral
exists. 
On the other hand, for $c \gg 1$ and $N \gg 1$, the combination in which 
$c$ and $N$ enter $P_N(c)$ depends explicitly on the tail of the 
underlying distribution $F$. This
indicates that somewhere in the regime of intermediate $c$, there is a
crossover in the $c$-dependence of $P_{N}(c)$ from a highly universal to a
less universal form.  

The result (\ref{GenSmallc}) has important
implications in the context of adaptive paths of evolutionary biology: 
Recalling that the expected number of accessible
paths between two genotypes which are $N$ mutations apart is given by 
$N! P_{N+1}$, we see in the presence of an arbitrarily small drift
this quantity \textit{increases} with $N$ as $c N$. Thus even a weak
systematic fitness gradient dramatically increases the
accessibility of mutational pathways in the direction of increasing
fitness.

\section*{Acknowledgements}

This work was supported by DFG within 
SFB 680 \textit{Molecular basis of evolutionary innovations} and 
the Bonn-Cologne Graduate School for Physics and Astronomy.

\section*{APPENDIX I - Computation of $I_n$ for the Gaussian distribution}
\label{AppendixI}

We begin by computing the saddle point $\tilde{y}$ defined by $\dd_y
g\left(\tilde{y}\right) = 0$, where the function $g(y)$ is given in 
(\ref{g}). The saddle point satisfies
\begin{eqnarray}
-2\tilde{y} +
\left(n-2\right)\frac{e^{-\frac{\tilde{y}^2}{2}}}{\int_{-\infty}^{\tilde{y}}
  \dd {y'} e^{-\frac{{y'}^2}{2}}} =
0. \label{saddle_point_equation}\end{eqnarray} 
For large $n$ this can only be solved by $\tilde y\gg1$, which implies that
$\int_{-\infty}^{\tilde{y}} \dd{y'} e^{-\frac{{y'}^2}{2}} \approx
\sqrt{2\pi}$ and reduces (\ref{saddle_point_equation}) to 
\begin{eqnarray}
\label{saddle_point_eq_approx}
\frac{\sqrt{8\pi}\ \tilde{y}}{n-2} = e^{\frac{-\tilde{y}^2}{2}}.
\end{eqnarray}
By taking the square on both sides of (\ref{saddle_point_eq_approx})
one finds that the solution is given in terms
of the Lambert W-function $\textrm{W}\left(z\right)$
\cite{Knuth,Wolfram} as
\begin{eqnarray}
\label{saddle_point}
 \tilde{y} = \sqrt{\textrm{W}\left(\frac{\left(n-2\right)^2}{8\pi}\right)}
\end{eqnarray}
(recall that $\textrm{W}\left(z\right)$ is defined implicitly through
$\textrm{W}(z) e^{\textrm{W}(z)} = z$). Using (\ref{saddle_point_eq_approx})
the function $g$ and its second derivative at the saddle point 
take the form
\begin{equation}
\label{g1}
g(\tilde y) \approx - {\tilde y}^2 - 2
\end{equation}
and 
\begin{equation}
\label{g2}
 \dd_y^2 g (\tilde y) \approx -2 (1 + {\tilde y}^2).
\end{equation}  
It follows that
\begin{equation}
\label{In_saddle}
I_n \approx \frac{n\left(n-1\right)}{4\pi} \sqrt{\frac{-2\pi}{\dd_y^2
    g\left(\tilde{y}\right)}}\; e^{g\left(\tilde{y}\right)}
\approx \frac{n\left(n-1\right)}{4\pi e^2} \sqrt{\frac{\pi}{1 +
    {\tilde y}^2}} \; e^{-{\tilde y}^2}. 
\end{equation}
Using once more (\ref{saddle_point_eq_approx}) to replace $e^{-{\tilde
    y}^2}$ we obtain
\begin{equation}
\label{In_saddle2}
I_n \approx  \frac{2 \sqrt{\pi}}{e^2} \frac{n(n-1)}{(n-2)^2}
\frac{{\tilde y}^2}{\sqrt{1 + {\tilde y}^2}} \to \frac{2
  \sqrt{\pi}}{e^2} {\tilde y} \approx  \frac{2 \sqrt{\pi}}{e^2} \sqrt{\ln
\left( \frac{n^2}{8 \pi} \right)}
\end{equation}
for large $n$, where we have used the expansion \cite{Knuth}
$\textrm{W}\left(z\right) \approx \textrm{ln}\left(z\right) -
\textrm{ln}\left(\textrm{ln}\left(z\right)\right)$ to evaluate 
(\ref{saddle_point}). This expansion also shows that the leading
corrections to the asymptotic expression (\ref{In_saddle2}) are of order
$\ln(\ln(n^2/8 \pi))/\ln(n^2/8 \pi)$, which accounts for the relatively large
deviations from the numerical results seen in figure \ref{GaussFigure}.

\section*{APPENDIX II - Generalized Gaussian distributions}
\label{AppendixII}

Here we consider probability densities of the form
\begin{equation}
\label{beta}
f(y) = C_\beta  e^{-|x|^\beta}
\end{equation}
with $\beta > 0$ and  
$
C_\beta = [2\Gamma(1+\frac{1}{\beta})]^{-1}.
$
We want to evaluate the integral (\ref{beta_integral})
in the saddle point approximation. Introducing the function 
\begin{equation}
\label{gbeta}
g(y) = - 2 y^\beta + (n-2) \ln \left( C_\beta \int_{-\infty}^y
\mathrm{d}y' e^{-|y'|^\beta} \right),
\end{equation}
the saddle point equation $\dd_y g\left(y\right) = 0$ reads, for large $n$, 
\begin{eqnarray}
\label{betasaddle}
 \frac{2 \beta C_\beta}{n-2}\tilde{y}^{\beta-1}
 = e^{-\tilde{y}^{\beta}}. 
\end{eqnarray}
The solution can again be expressed in terms of the Lambert-W function. Defining
$\eta:=1-\frac{1}{\beta}$ we find 
\begin{eqnarray}
 \tilde{y} \approx \left(\eta \textrm{W}\left(\eta^{-1}
     \left(\frac{n-2}{2\beta C_\beta}\right)^{\eta^{-1}}\right)\right)^{\frac{1}{\beta}}.   
\end{eqnarray}
Note that this expression is valid both for $\beta > 1$ ($\eta > 0$)
and for $\beta < 1$ ($\eta < 0$), but in the latter case the second
real branch of $\textrm{W}(z)$ has to be used \cite{Knuth}. Using the asymptotics of $W(z)$ we obtain
\begin{equation}
\label{Wbeta}
\tilde y \approx \left( \ln n - \eta \ln(\eta^{-1} \ln n)
\right)^{1/\beta} \approx (\ln n)^{1/\beta}
\end{equation}
for large $n$.

With the help
of (\ref{betasaddle}) the function $g$ and its second derivative at
the saddle point become
\begin{equation}
\label{gbeta2}
g(\tilde y) \approx - 2 {\tilde y}^\beta - 2 \approx - 2 {\tilde y}^\beta
\end{equation}
and
\begin{equation}
\label{gbeta3}
\dd_y^2g\left(\tilde{y}\right) \approx - 2 \beta (\beta-1) {\tilde
  y}^{\beta - 2} - 2 \beta^2 {\tilde y}^{2 \beta - 2} \approx - 2
\beta^2 {\tilde y}^{2 \beta - 2}
\end{equation}
for large $\tilde y$. Thus, using (\ref{betasaddle}), we see that
$e^{g(\tilde y)} \approx e^{- {2 \tilde y}^\beta} \sim {\tilde y}^{2 \beta  - 2}/n^2$, and therefore (ignoring all constant prefactors)                                                         
\begin{eqnarray}
I_n \sim n^2 \sqrt{\frac{-2\pi}{\dd_y^2
    g\left(\tilde{y}\right)}}e^{g\left(\tilde{y}\right)} \sim {\tilde
  y}^{\beta - 1} \sim (\ln n)^{1 - 1/\beta}.
\end{eqnarray}

\section*{Appendix III - Proof of an expansion}
\label{AppendixIII}
In this appendix, we will provide the details on the expansion of
$\int_{-\infty}^{\infty}\dd x f^{2}(x)$ into the terms on the right
hand side of (\ref{ChainOfIntegrals}). The starting  point is the relation
\begin{eqnarray}\label{IntByParts}
  F^{n}(x)\int_{-\infty}^{x}\dd y
  f^{2}(y)=&n\int_{-\infty}^{x}\dd y
  f(y)F^{n-1}(y)\int_{-\infty}^{y}\dd z f^{2}(z)\nonumber\\ 
  &+ \int_{-\infty}^{x}\dd z f^2(z)F^{n}(z), 
\end{eqnarray}
which can be proved by applying integration by parts to the first term
on the right hand side. 
With the identities $F^{n}(\infty)=1$ and
$F^{n}(-\infty)=0$, one obtains from (\ref{IntByParts}) 
\begin{eqnarray}\label{helper}
\fl  \frac{1}{n!}\int_{-\infty}^{\infty}\dd z
  f^2(z)&=&\left.\frac{F^{n}(x)}{n!}\int_{-\infty}^{x}\dd z
  f^{2}(z)\right|_{x=-\infty}^{\infty} \nonumber\\ 
\fl   &=&\frac{F^{n}(\infty)}{n!}\int_{-\infty}^{\infty}\dd z
  f^{2}(z)-\frac{F^{n}(-\infty)}{n!}\int_{-\infty}^{-\infty}\dd x f^{2}(x)\\ 
\fl  &=&\frac{1}{n!}\int_{-\infty}^{\infty}\dd z
  f^{2}(z)F^{n}(z) +\frac{1}{(n-1)!}\int_{-\infty}^{\infty}\dd z
  f(z)F^{n-1}(z)\int_{-\infty}^{z}\dd z' f^{2}(z'). \nonumber
\end{eqnarray}
The first term of the sum is already identical to the first term in
(\ref{ChainOfIntegrals}) for $n\equiv N-2$. Using
(\ref{IntByParts}) on the inner of the two integrals of the second term
of the sum above, one obtains
\begin{eqnarray*}
 \fl \int_{-\infty}^{\infty}\dd z
  f(z)F^{n-1}(z)\int_{-\infty}^{z}\dd z'
  f^2(z')&=&\int_{-\infty}^{\infty}\dd
  zf(z)\int_{-\infty}^{z}\dd z'f^{2}(z')F^{n-1}(z')\\
   \fl &+&(n-1)\int_{-\infty}^{\infty}\dd z
  f(z)\int_{-\infty}^{z}\dd z' f(z')F^{n-2}(z')\int_{-\infty}^{z'}\dd
  z''f^{2}(z''). 
\end{eqnarray*}
Dividing by $(n-1)!$ and putting this back into (\ref{helper}) with
$n=N-2$, one sees that now the first \emph{two} terms of the 
sum agree with (\ref{ChainOfIntegrals}). By repeating this
procedure on the terms that do not yet match and noting that finally
$F^{0}(z)=1$, one has expanded $\int_{-\infty}^{\infty}\dd x f^{2}(x)$
into the RHS of (\ref{ChainOfIntegrals}), which concludes the
proof of (\ref{GenSmallc}).

\end{document}